\documentclass[preprint,epsfig,onecolumn,floats, showkeys]{revtex4}
\usepackage{makeidx}
\usepackage{amsmath}
\usepackage{amssymb}
\usepackage{graphicx}
\usepackage{epsfig}

\begin{document}

\title{Quantum Discrete Fourier Transform with Classical Output for Signal
Processing}
\author{Chao-Yang Pang}
\email{cyp_900@hotmail.com}
\email{cypang@sicnu.edu.cn}
\affiliation{College of Mathematics and Software Science, Sichuan Normal University,
Chengdu 610066, P.R. China}
\author{Ben-Qiong Hu}
\affiliation{College of Information Management, Chengdu University of Technology, P.R.
China}

\begin{abstract}
Discrete Fourier transform (DFT) is the base of modern signal or information
processing. 1-Dimensional fast Fourier transform (1D FFT) and 2D FFT have
time complexity $O(NlogN)$ and $O(N^{2}logN)$ respectively. Quantum 1D and
2D DFT algorithms with classical output (1D QDFT and 2D QDFT) are presented
in this paper. And quantum algorithm for convolution estimation is also
presented in this paper. Compared with FFT, QDFT has two advantages at
least. One of advantages is that 1D and 2D QDFT has time complexity $O(\sqrt{%
N})$ and $O(N)$ respectively. The other advantage is that QDFT can process
very long signal sequence at a time. QDFT and quantum convolution
demonstrate that quantum signal processing with classical output is possible.
\end{abstract}

\maketitle

\section{Introduction}

\subsection{Introduction of Discrete Fourier transform (DFT)}

DFT is the base of modern signal and information processing. No DFT, no
modern signal and information processing \cite{Hu}.

Let

\begin{equation}
W_{N}=\frac{1}{\sqrt{N}}\left[
\begin{array}{ccccc}
\omega ^{0} & \omega ^{0} & \omega ^{0} & \cdots & \omega ^{0} \\
\omega ^{0} & \omega ^{1} & \omega ^{2} & \cdots & \omega ^{N-1} \\
\omega ^{0} & \omega ^{2} & \omega ^{4} & \cdots & \omega ^{2(N-1)} \\
\vdots & \vdots & \vdots & \vdots & \vdots \\
\omega ^{0} & \omega ^{N-1} & \omega ^{2(N-1)} & \cdots & \omega
^{(N-1)(N-1)}%
\end{array}%
\right] =\left(
\begin{array}{c}
\overrightarrow{W_{0}} \\
\overrightarrow{W_{1}} \\
\vdots \\
\overrightarrow{W_{N-1}}%
\end{array}%
\right)  \label{eqDFTMatrix}
\end{equation}

, where $\omega =e^{-i\frac{2\pi }{N}}$ and $\overrightarrow{W_{0}}$, $%
\overrightarrow{W_{1}}$, $\cdots $, $\overrightarrow{W_{N-1}}$ is the line
vector of the matrix. The matrix $W_{N}$ defined in Eq.\ref{eqDFTMatrix} is
called \textbf{Fourier transform matrix} \cite{Hu}.

\textbf{1-Dimensional DFT (1D DFT) }\cite{Hu} is defined as

\begin{equation}
\overrightarrow{c}=W_{N}\overrightarrow{x}=\left(
\begin{array}{c}
\overrightarrow{W_{0}}\bullet \overset{\rightarrow }{x} \\
\overrightarrow{W_{1}}\bullet \overset{\rightarrow }{x} \\
\vdots \\
\overrightarrow{W_{N-1}}\bullet \overset{\rightarrow }{x}%
\end{array}%
\right)  \label{eq1DDFT}
\end{equation}

, where vector $\overrightarrow{x}=(x_{0},x_{1},x_{2},\cdots ,x_{N-1})^{T}$,
$\overrightarrow{c}=(c_{0},c_{1},c_{2},\cdots ,c_{N-1})^{T}$ ($T$ denotes
the transpose of vector), and \textquotedblleft $\bullet $%
\textquotedblright\ denotes the inner product between two vectors. Each
component of vector $\overrightarrow{c}$ is called \textbf{Fourier
coefficient}.

For any input matrix $F=[f_{ij}]_{N\times N}$ , \textbf{2D DFT }\cite{Hu} is
defined as
\begin{equation}
C=W_{N}FW_{N}  \label{eq2DDFT}
\end{equation}%
, where $C$ denotes the matrix of DFT coefficients.

DFT has two important properties, that will be applied to design quantum
algorithm in this paper. One property is that DFT is energy conservation
transform (i.e., $||\overrightarrow{x}||^{2}=||\overrightarrow{c}||^{2}$, or
$\sum (x_{i})^{2}=\sum |c_{i}|^{2}$). The other property is that, typical
data sequence, such as digital image, has high redundance, many of its DFT
coefficients have values close to zero, and these coefficients can be
discarded without seriously affecting the estimated value of $||%
\overrightarrow{x}||^{2}$ (e.g., without seriously affecting the quality of
the restored image). Therefore, inverse DFT acting on the few big
coefficients retained can restore the original data approximatively. Fig.\ref%
{figDFT} shows the properties of DFT.

\begin{figure}[tbh]
\epsfig{file=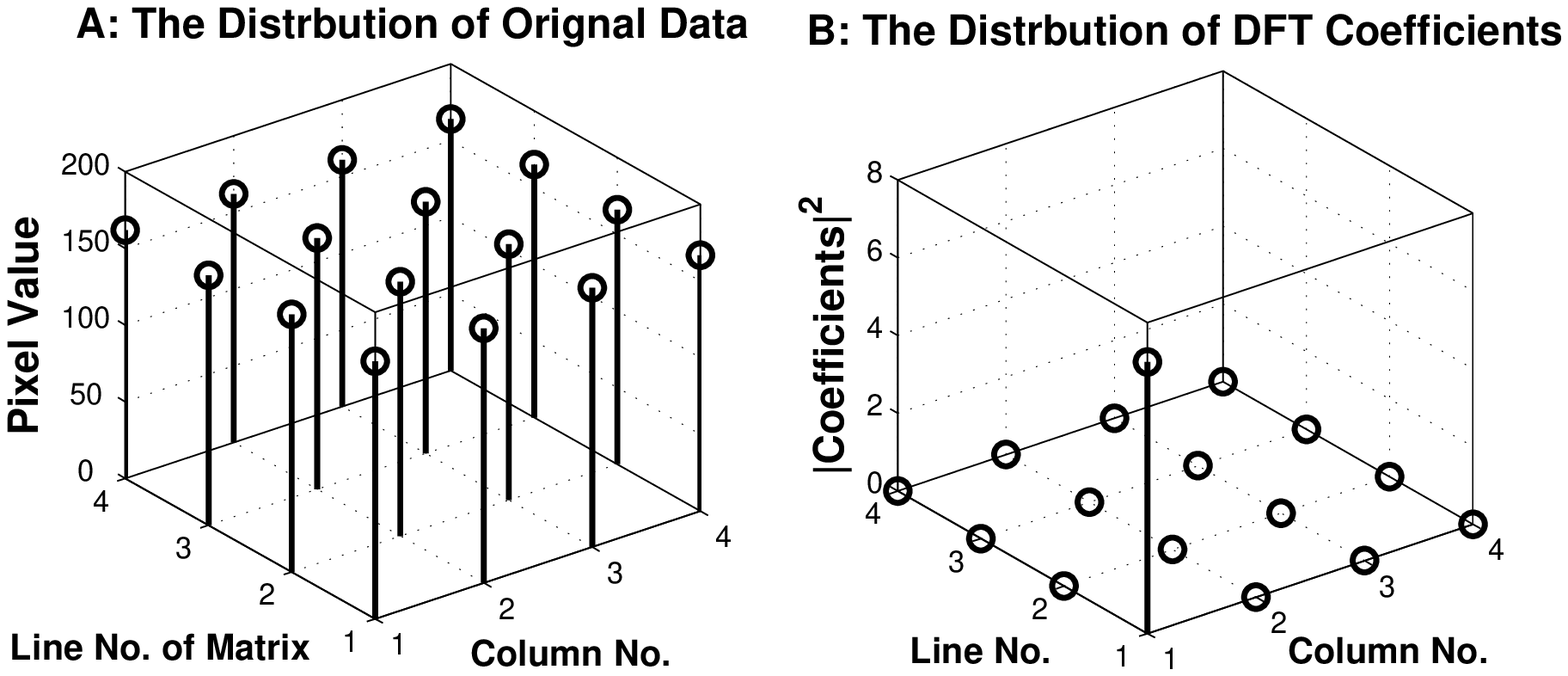,width=10cm,} \caption{\textbf{The Illustration
of the Two Properties of DFT that Applied to Design Quantum
Algorithm in This Paper:} fig.A denotes the original data
of the top-left image block with size 4$\times $4 of image Lena \protect\cite%
{QVQ2}. Fig.B denotes the 2D DFT coefficients of the image block. DFT has
two properties. One property is that DFT is energy conservation transform.
The other property is that, only few DFT coefficients is \emph{not} close to
zero, and the inverse DFT acting on these coefficients can restore the
original data approximatively. The two properties are applied to design
quantum DFT algorithm in this paper.}
\label{figDFT}
\end{figure}

Fast Fourier transforms (FFT) with time complexity $O(NlogN)$ and $%
O(N^{2}logN)$ for 1D and 2D DFT respectively were presented at 1965 \cite{Hu}%
. And the desire for designing more fast DFT algorithm is still very strong.
However, there is no more fast algorithm is presented up till now. Classical
computation can not improve the efficiency of DFT any more maybe, and the
principle of classical computation becomes the bottleneck of designing fast
DFT algorithm maybe. Can quantum computation be applied to design more fast
DFT algorithm with classical output to break the bottleneck?

\subsection{Introduction of Quantum Computation}

\textbf{Introduction of Quantum Fourier Transform (QFT):} QFT \cite%
{Shor,Galindo} on an orthonormal basis $|0\rangle $, $|1\rangle $, $\cdots $%
, $|N-1\rangle $ is defined to be a linear operator $QFT|j\rangle =\frac{1}{%
\sqrt{N}}\overset{N-1}{\underset{k=0}{\sum }}e^{2\pi ijk/N}|k\rangle $. QFT
is the key of the famous Shor's order-finding and factoring algorithm \cite%
{Shor}. QFT has time complexity $O(log^{2}N)$. However, QFT is not suitable
to signal and information processing because the result of DFT defined in Eq.%
\ref{eq1DDFT} (or Eq.\ref{eq2DDFT}) can \emph{not} be generated and measured
out by this QFT \cite{Nielsen,QCZhao}.

\textbf{Introduction of} \textbf{Grover's Algorithm:} Grover's algorithm
\cite{Grover} solves the problem of searching for an element with a unique
index $i_{0}$ in a list of $N$ unsorted elements, similar to searching a
database like a telephone directory when we know the number but not the
person's name \cite{Galindo}. Grover's algorithm has time complexity $O(%
\sqrt{N})$ \cite{Galindo}. Long proposes a modified Grover's algorithm, that
has the probability of success 100\% even for the case that the number of
elements is very small \cite{GLLong}. Boyer, Brassard, H$\phi $yer, and Tap
present the modified Grover's algorithm named BBHT algorithm in this paper
for the case that the number of solutions is unknown \cite{BBHT}. BBHT
algorithm is a very smart algorithm because it saves many quantum circuits.

\textbf{Introduction of Quantum Loading Scheme }$U_{L}$ \cite{QLS}\textbf{: }%
Grover's algorithm has the function that find the index $i_{0}$ of a special
database record $record_{i_{0}}$ from the index superposition of state $%
\frac{1}{\sqrt{N}}({\sum\limits_{i=0}^{N-1}{{{\left\vert {i}\right\rangle )}}%
}}$. And the record $record_{i_{0}}$\ is the genuine answer wanted by us.
However, the corresponding record $record_{i_{0}}$ can not be measured out
unless the 1-1mapping relationship between index $i$ and the corresponding
record $record_{i}$ is bound in the entangled state $\frac{1}{\sqrt{N}}({%
\sum\limits_{i=0}^{N-1}{{{\left\vert {i}\right\rangle {{{\left\vert
record_{i}\right\rangle }}})}}}}$. That is, we need a unitary$\ $operation $%
U_{L}$ to load all records that are stored in a classical database into
quantum state. The function of unitary$\ $operation $U_{L}$ can be described
as
\begin{equation}
\frac{1}{\sqrt{N}}({\sum\limits_{i=0}^{N-1}{{{\left\vert {i}\right\rangle
\left\vert 0\right\rangle )\left\vert {ancilla}\right\rangle }}}}\overset{%
U_{L}}{{\rightarrow }}\frac{1}{\sqrt{N}}({\sum\limits_{i=0}^{N-1}{{{%
\left\vert {i}\right\rangle \left\vert record_{i}\right\rangle )\left\vert {%
ancilla}\right\rangle }}}}  \label{eqUL}
\end{equation}

, where ancillary state ${{{{\left\vert {ancilla}\right\rangle }}}}$ is
known.

Pang proposes a design method of the unitary operation $U_{L}$, that has
time complexity $O(logN)$ (unit time: phase transform and flipping the
qubits of registers) \cite{QLS}. Operator $U_{L}$ is so fast that its
running time can be ignored when analyzing the time complexity of a
algorithm.

\textbf{Introduction of Quantum Search Algorithm with Complex Computation
(i.e., the Method of Rotation at Subspace) \cite{QVQ1,QDCT,QVQ2,QIC}:}

Grover's algorithm can find a database record according to the given index.
However, database search is complex in general. E.g., police often hopes to
find a mug shot from the database in which many sample photos are stored by
the method of matching every sample photo and the photo captured by the
vidicon at the entrance of airport real-time. Grover's algorithm is invalid
for this kind of search case because the coupling between search and other
computation (e.g., image matching) is required at this case. Pang et.al.
presents a quantum method named "\textbf{rotation at subspace}" \cite%
{QVQ1,QDCT,QVQ2,QIC} to generalize Grover's algorithm to the search case
with arbitrary complex computation, that is derived from the research of
quantum image compression \cite{QIC}. The method of rotation at subspace is
described as following briefly:

First, All input datum are stored in classical memory as database records.
Assume that total number of records is $N$. All these records can be loaded
into a superposition of state using quantum loading scheme $U_{L}$.

Second, construct the \textbf{general Grover iteration (GGI)} $G_{general}$
as

\begin{equation}
G_{general}=(2|\xi \rangle \langle \xi |-I)(U_{L})^{\dag }(O_{c})^{\dag
}O_{f}O_{c}U_{L}  \label{eqGeneralGroverIteration}
\end{equation}

, where $O_{c}$ denotes computation oracle such as image matching, $f\ $%
denotes the judge function (i.e., if the output of $O_{c}$ satisfies some
conditions, let $f=1$, else $f=0$),$\ O_{f}$ is the oracle of the judge
function, and $|\xi \rangle =\frac{1}{\sqrt{N}}\sum\limits_{i=0}^{N-1}|i%
\rangle $.

Third, similar to Grover's algorithm, let unitary operation $G_{general}$
act on initial state $\frac{1}{\sqrt{N}}(\underset{i=0}{\overset{N-1}{\sum }}%
{\left\vert {i}\right\rangle }_{{registers1}}){\left\vert 0\right\rangle }%
_{registers2}$ $O(\sqrt{N})$ times, we will find the optimal solution.

\section{1-Dimensional Quantum DFT (1D QDFT)}

\textbf{First, construct the following data structure (DS) and unitary
operations :}

\textbf{DS1.} Save DFT matrix $W$ defined in Eq.\ref{eqDFTMatrix} in
classical memory as a database. And each line vector $\overrightarrow{W_{i}}$
of the matrix is regard as a record, and these records have indices $%
0,1,...,N-1$.

\textbf{DS2.} Construct six registers that have data format
\begin{equation*}
|\alpha \rangle _{register1}|\beta \rangle _{register2}|\overrightarrow{x}%
\rangle _{register3}|i\rangle _{register4}|\overrightarrow{W_{i}}\rangle
_{register5}|(\overrightarrow{W_{i}}\bullet \overrightarrow{x})^{2}\rangle
_{register6}
\end{equation*}%
. That is, 1st, 2nd, 3rd, 4th, 5th, and 6th register are used to save input
parameter $\alpha $, input parameter $\beta $, input vector $\overrightarrow{%
x}$, index $i$, line vector $\overrightarrow{W_{i}}$, and squared inner
product respectively.

\textbf{DS3.} Design oracle $O_{inner}$ to compute the squared inner product
between vector $\overrightarrow{W_{i}}$ and $\overrightarrow{x}$, i.e.,

\begin{equation*}
|\alpha \rangle |\beta \rangle |\overrightarrow{x}\rangle |i\rangle |%
\overrightarrow{W_{i}}\rangle |0\rangle \overset{O_{inner}}{\rightarrow }%
|\alpha \rangle |\beta \rangle |\overrightarrow{x}\rangle |i\rangle |%
\overrightarrow{W_{i}}\rangle |(\overrightarrow{W_{i}}\bullet
\overrightarrow{x})^{2}\rangle
\end{equation*}

\ \ \ \ \ We can design very simple parallel circuit to calculate squared
inner product, and has time complexity $2t_{m}+\lceil log_{2}N\rceil t_{a}$,
where $t_{m}$ and $t_{a}$ denote the unit running time of multiplication and
addition respectively. Because addition is more fast than multiplication ($%
50t_{a}<t_{m}$ in general) and $N<2^{50}$ in general, the running time of $%
O_{inner}$ can be regarded as few times of multiplication (i.e., $O(1)t_{m}$%
).

\textbf{DS4.} Design oracle $O_{f}$:

\begin{equation*}
|\alpha \rangle |\beta \rangle |\overrightarrow{x}\rangle |i\rangle |%
\overrightarrow{W_{i}}\rangle |(\overrightarrow{W_{i}}\bullet
\overrightarrow{x})^{2}\rangle \overset{O_{f}}{\rightarrow }%
(-1)^{f(i)}|\alpha \rangle |\beta \rangle |\overrightarrow{x}\rangle
|i\rangle |\overrightarrow{W_{i}}\rangle |(\overrightarrow{W_{i}}\bullet
\overrightarrow{x})^{2}\rangle
\end{equation*}

, where $f(i)=\left\{
\begin{tabular}{ccc}
$1$ & $if$ & $\alpha \leq (\overrightarrow{W_{i}}\bullet \overrightarrow{x}%
)^{2}\leq \beta $ \\
$0$ & \multicolumn{2}{c}{$otherwise$}%
\end{tabular}%
\right. $.

Oracle $O_{f}$ is used to mark the DFT coefficients $\overrightarrow{W_{i}}%
\bullet \overrightarrow{x}$ that satisfy the condition $\alpha \leq (%
\overrightarrow{W_{i}}\bullet \overrightarrow{x})^{2}\leq \beta $.

\textbf{DS5.} Define 1D\ QDFT\ iteration $G_{1DQDFT}$:

\ \ \ \ According to Eq.\ref{eqGeneralGroverIteration},\ 1D\ QDFT\
iteration\ $G_{1DQDFT}$\ is

\begin{equation}
G_{1DQDFT}=(2|\xi \rangle \langle \xi |-I)(U_{L})^{-1}(O_{inner})^{\dagger
}O_{f}O_{inner}U_{L}  \label{eq1DQDFT}
\end{equation}

\textbf{Second, design the following subroutine 1 to find a coefficient }$%
\overrightarrow{W_{i_{0}}}\bullet \overrightarrow{x}$\textbf{\ that
satisfies the condition }$\alpha \leq (\overrightarrow{W_{i_{0}}}\bullet
\overrightarrow{x})^{2}\leq \beta $\textbf{:}

\textbf{subroutine 1:}

\textbf{Step1.} Initialize $m=1$ and set $\lambda =6/5$. (Any value of $%
\lambda $ strictly between 1 and 4/3 would do.)

\textbf{Step2.} Choose $j$ uniformly at random among the nonnegative
integers smaller than $m$.

\textbf{Step3.} Apply $j$ iterations of $G_{1DQDFT}$\ acting on initial
state $|\psi _{0}\rangle =\frac{1}{\sqrt{N}}|\alpha \rangle |\beta \rangle |%
\overrightarrow{x}\rangle (\sum\limits_{i=0}^{N-1}|i\rangle )|0\rangle
|0\rangle $.

\textbf{Step4.} Observe the 4th register: let $i_{0}$ be the outcome.

\textbf{Step5.} Calculate value $\overrightarrow{W_{i_{0}}}\bullet
\overrightarrow{x}$ using classical computation. If $\alpha \leq (%
\overrightarrow{W_{i_{0}}}\bullet \overrightarrow{x})^{2}\leq \beta $,
preserve $i_{0}$, and exit.

\textbf{Step6.} Otherwise, set $m$ to $min(\lambda m,\sqrt{N})$ and go back
to step 2.

Subroutine 1 is similar to BBHT algorithm \cite{BBHT}, and the main
difference between them is that Grover iteration is replaced by 1D\ QDFT\
iteration $G_{1DQDFT}$ that realizes the coupling between quantum search and
the computation of inner product. Subroutine 1 has time complexity $O(\sqrt{%
\frac{N}{M}})$ \cite{BBHT}, where $M$ denotes the number of coefficients $%
\overrightarrow{W_{i}}\bullet \overrightarrow{x}$ that satisfy the condition
$\alpha \leq (\overrightarrow{W_{i}}\bullet \overrightarrow{x})^{2}\leq
\beta $ .

\textbf{Third, design the following 1D QDFT algorithm:}

\textbf{Step 0.} Let $\Delta E=||\overrightarrow{x}||^{2}=\overset{N--1}{%
\underset{i=0}{\sum }}(x_{i}\times x_{i})$, $\alpha =\frac{\Delta E}{N}$, $%
\beta =\Delta E$, $nS=0$. We can design a very simple parallel circuit to
calculate value$\ \Delta E$, and the parallel circuit has computation
complexity $O(1)$ (unit time : multiplication) approximately.

\textbf{Step 1.} Generate the initial state $|\psi _{0}\rangle =\frac{1}{%
\sqrt{N}}|\alpha \rangle |\beta \rangle |\overrightarrow{x}\rangle
(\sum\limits_{i=0}^{N-1}|i\rangle )|0\rangle |0\rangle $. This can be
achieved in $O(log_{2}N)$ steps using a $\lceil \log _{2}N\rceil -bit$
Hadamard transform, which is so fast that the running time can be ignored.

\textbf{Step 2.} while($\frac{\Delta E}{||\overrightarrow{x}||^{2}}\geq
\varepsilon $)

, where $\varepsilon $ is the given threshold.

\{

\ \ \ \textbf{Step 2.1:} Apply subroutine 1 to find a coefficient $c[i_{0}]=%
\overrightarrow{W_{i_{0}}}\bullet \overrightarrow{x}$ that satisfies the
condition $\alpha \leq (c[i_{0}])^{2}\leq \beta $;

\ \ \ \textbf{Step 2.2:} If the DFT coefficient $c[i_{0}]$ is the result
that is not be found previously by this algorithm, preserve it, and let $%
\Delta E=\Delta E-(c[i_{0}])^{2}$, $nS=nS+1$, $\alpha =\frac{\Delta E}{N-nS}$%
, and $\beta =\Delta E$, where $\alpha $ denotes the average residual energy
per DFT\ coefficient which is not be obtained still by this algorithm, $%
\beta $ denotes the total residual energy that is preserved by the DFT
coefficients which are not be obtained still by this algorithm, and $nS$
denotes the number of the coefficients which have been obtained by this
algorithm.

\}

It's the main idea of step 2.1 that apply subroutine 1 to find a coefficient
which energy is bigger than the average residual energy (i.e., value $\alpha
$) and smaller than the total residual energy (i.e., value $\beta $).
Because DFT is energy conversation transform (i.e., $\overset{N--1}{\underset%
{i=0}{\sum }}(x_{i})^{2}=\overset{N--1}{\underset{i=0}{\sum }}(c_{i})^{2}$),
the stop criterion $\frac{\Delta E}{||\overrightarrow{x}||^{2}}\geq
\varepsilon $ shows that the above algorithm will find all big coefficients
that preserve nearly all energy and information if threshold $\varepsilon $
is enough small. In addition, almost signal sequences have high redundance
and the main task of DFT is to find and retain big coefficients to eliminate
this redundance (see Fig.\ref{figDFT}). Therefore, 1D QDFT algorithm
realizes 1D DFT computation approximately.

1D QDFT has time complex $O(\sqrt{mN})$ approximately, where $m$ denotes the
number of the big coefficients. Because signal sequences have high
redundance in general, the case $m\ll N$ often happens (see Fig.\ref{figDFT}%
). Thus, we regard 1D QDFT has time complexity $O(\sqrt{N})$ (unit:
multiplication and addition), while classical 1D FFT has time complexity $%
O(Nlog_{2}N)$ (unit: multiplication and addition) and even parallel DFT
computation has time complexity $O(N)$. QDFT can process long signal
sequence at a time, while classical computer requires that long signal
sequence must be divded into many small sections (such as $8\times 8$ or $%
16\times 16$ section) on which DFT acts sequentially to evade the efficiency
bottleneck of loading data \cite{QLS}.

\section{2D QDFT}

2-D DFT (see Eq.\ref{eq2DDFT}) is a separable linear transformation. That
is, the result of 2-D DFT may be obtained by first taking transforms along
the columns of $F$ and then along the rows of that result \cite{Hu}. That
is, $C=W_{N}FW_{N}=(W_{N}F)W_{N}$. We define
\begin{equation*}
G=W_{N}F=(\overrightarrow{W_{i}}\bullet \overrightarrow{f_{j}})_{N\times N}
\end{equation*}%
, where $\overrightarrow{W_{i}}$ is the line vector of matrix $W_{N}$ and $%
\overrightarrow{f_{j}}$ is the column vector of input matrix $F$.

The main task of 2D QDFT is to calculate out matrix $G$ and matrix $GW_{N}$
(i.e., $C$). 2D QDT is described as following:

\textbf{First, }Construct the following data structures (DS) and unitary
operations:

\textbf{DS1: }Store all of line vector $\overrightarrow{W_{i}}$ in a
database, and each index of line vector is denoted by $i$. Store all of
column vector $\overrightarrow{f_{j}}$ of matrix $F$ in the database, and
each index of column vector is denoted by $j$.

\textbf{DS2: }Construct seven registers that has data format $|\alpha
\rangle |\beta \rangle |i\rangle |j\rangle |\overrightarrow{W_{i}}\rangle |%
\overrightarrow{f_{j}}\rangle |(\overrightarrow{W_{i}}\bullet
\overrightarrow{f_{j}})^{2}\rangle $.

\textbf{DS3:} Design two loading operation $U_{L1}$ and $U_{L2}$ according
to Eq.\ref{eqUL} and ref.\cite{QLS}, where $U_{L1}$ and $U_{L2}$ are applied
to load vector $\overrightarrow{W_{i}}$ and $\overset{\rightarrow }{f_{j}}$
into registers from classical database respectively. That is,

\begin{equation*}
\left\{
\begin{tabular}{c}
$|\alpha \rangle |\beta \rangle |i\rangle |j\rangle |0\rangle |0\rangle
|0\rangle \overset{U_{L1}}{\rightarrow }|\alpha \rangle |\beta \rangle
|i\rangle |j\rangle |\overrightarrow{W_{i}}\rangle |0\rangle |0\rangle $ \\
$|\alpha \rangle |\beta \rangle |i\rangle |j\rangle |0\rangle |0\rangle
|0\rangle \overset{U_{L2}}{\rightarrow }|\alpha \rangle |\beta \rangle
|i\rangle |j\rangle |0\rangle |\overrightarrow{f_{j}}\rangle |0\rangle $%
\end{tabular}%
\right.
\end{equation*}%
\

\ \textbf{DS4:} Design oracle $B_{inner}$ to calculate squared inner
product. That is,

\begin{equation*}
|\alpha \rangle |\beta \rangle |i\rangle |j\rangle |\overrightarrow{W_{i}}%
\rangle |\overrightarrow{f_{j}}\rangle |0\rangle \overset{B_{inner}}{%
\rightarrow }|\alpha \rangle |\beta \rangle |i\rangle |j\rangle |%
\overrightarrow{W_{i}}\rangle |\overrightarrow{f_{j}}\rangle |(%
\overrightarrow{W_{i}}\bullet \overrightarrow{f_{j}})^{2}\rangle
\end{equation*}

\ \textbf{DS5:} Design oracle $O_{f}^{\prime }$ :

\begin{equation*}
|\alpha \rangle |\beta \rangle |i\rangle |j\rangle |\overrightarrow{W_{i}}%
\rangle |\overrightarrow{f_{j}}\rangle |(\overrightarrow{W_{i}}\bullet
\overrightarrow{f_{j}})^{2}\rangle \overset{O_{f}^{\prime }}{\rightarrow }%
(-1)^{f^{\prime }(i,j)}|\alpha \rangle |\beta \rangle |i\rangle |j\rangle |%
\overrightarrow{W_{i}}\rangle |\overrightarrow{f_{j}}\rangle |(%
\overrightarrow{W_{i}}\bullet \overrightarrow{f_{j}})^{2}\rangle
\end{equation*}

, where $f^{\prime }(i,j)=\left\{
\begin{tabular}{ccc}
$1$ & $if$ & $\alpha \leq (\overrightarrow{W_{i}}\bullet \overrightarrow{%
f_{j}})^{2}\leq \beta $ \\
$0$ & \multicolumn{2}{c}{$otherwise$}%
\end{tabular}%
\right. $.

\ \textbf{DS6:} Design 2D QDFT iteration $G_{2DQDFT}$:

\begin{equation*}
G_{2DQDFT}=(2|\xi ^{\prime }\rangle \langle \xi ^{\prime
}|-I)(U_{L1})^{\dagger }(U_{L2})^{\dagger }(B_{inner})^{\dagger
}O_{f}^{\prime }B_{inner}U_{L2}U_{L1}
\end{equation*}

, where $|\xi ^{\prime }\rangle =\frac{1}{N}\sum\limits_{i=0}^{N-1}\sum%
\limits_{j=0}^{N-1}\underbrace{|i\rangle |j\rangle }$.

\textbf{Second,} Generate the initial state $|\Psi _{0}\rangle =\frac{1}{N}%
\sum\limits_{i=0}^{N-1}\sum\limits_{j=0}^{N-1}|\alpha \rangle |\beta \rangle
\underbrace{|i\rangle |j\rangle }|0\rangle |0\rangle |0\rangle $. And apply
the method of 1D QDFT, we will calculate out all elements of matrix $G$.
Apply the same method again, we can calculate out all elements of matrix $C$.

2D QDFT has time complexity $O(N)$, while classical 2D FFT has time
complexity $O(N^{2}log_{2}N)$.

\section{Using 1D QDFT to Calculate Convolution}

For the given two N-dimensional vectors $\overrightarrow{u}%
=(u_{0},u_{1},...,u_{N-1})$ and $\overrightarrow{v}%
=(v_{0},v_{1},...,v_{N-1}) $, let $w_{k}=\underset{j=0}{\overset{N-1}{\sum }}%
u_{j}v_{(k-j)modN}$, where $0\leq k<N$. Then vector $\overrightarrow{w}%
=(w_{0},w_{1},...,w_{N-1})$ is called the \textbf{periodic convolution of
vectors }$\overrightarrow{u}$\textbf{\ and }$\overrightarrow{v}$. We often
denote the convolution by symbol $\overrightarrow{w}=\overrightarrow{u}\ast
\overrightarrow{v}$. If $\overrightarrow{u}$ and $\overrightarrow{v}$ have
different dimensions, add zero components to each vector such that both
vectors have the same dimension. The periodic convolution of the new vectors
can be regarded as the result. Algebraically, convolution is the same
operation as multiplying the polynomials whose coefficients are the elements
of $\overrightarrow{u}$ and $\overrightarrow{v}$ \cite{Hu}.

Holding the view of signal processing, we can regard vector $\overrightarrow{%
u}$ as input signal sequence and vector $\overrightarrow{v}$ is the function
of a physical apparatus. After vector $\overrightarrow{u}$ passing through
the apparatus, convolution $\overrightarrow{w}$ is came out. Furthermore,
the output sequence $\overrightarrow{w}$ becomes smoother than the input
sequence $\overrightarrow{u}$ in general. Therefore, convolution is very
important for signal or information processing \cite{Hu}.

The convolution theorem \cite{Hu} says, roughly, that convolving two
sequences is the same as multiplying their Fourier transforms. That is \cite%
{Hu},
\begin{equation}
DFT(\overrightarrow{u}\ast \overrightarrow{v})=DFT(\overrightarrow{u})DFT(%
\overrightarrow{v})  \label{eqConvolution}
\end{equation}

In general, the big coefficients of $DFT(\overrightarrow{v})$ are few and
other coefficients have little contribution to the result. That is, we can
calculate only big coefficients of $DFT(\overrightarrow{u})$ and $DFT(%
\overrightarrow{v})$ and discard small coefficients to speedup the
computation of convolution. According to this property, \textbf{quantum
algorithm for convolution estimation} is described as following:

\textbf{Step1.} Calculate $DFT(\overrightarrow{u})$ and $DFT(\overrightarrow{%
v})$ using the 1D QDFT.

\textbf{Step2.} Calculate $DFT(\overrightarrow{u}\ast \overrightarrow{v})$
according to Eq.\ref{eqConvolution}.

\textbf{Step3.} Calculate the inverse of $DFT(\overrightarrow{u}\ast
\overrightarrow{v})$ to obtain convolution $\overrightarrow{u}\ast
\overrightarrow{v}$.

It should be noticed that the efficiency of the above algorithm depends on
the redundancy (or smooth property) of the signal sequences. The more higher
the redundancy is, the more fast the quantum algorithm runs. Fortunately,
many digital signal sequences have high redundancy, that is just the
existence reason of modern signal processing technique.

\begin{acknowledgments}
The first author thanks his teacher prof. G.-C. Guo and the Key Lab.
of Quantum Information, USTC, P. R. of China for the author is
brought up from the lab. The first author thanks prof. Z. F. Han's
for he encouraging and helping the author up till now. The first
author thanks Dr. Z.-W. Zhou's help and useful discussion from 2004
to 2006. The first author thanks prof. V. N. Gorbachev who is at
St.-Petersburg State Uni. of Aerospace Instrumentation for the
discussion about quantum discrete cosine transform  at 2006. The
first author thanks assistant prof. Xudong Huang who is at Harvard
Uni. for his interest at the author's research topic of quantum
image compression and the discussion. The author thanks Dr. J.-M.
Cai for his help and discussion. The author thanks prof. J. Zhang
and prof. Z.-L. Pu who are at Sichuan Normal Uni. for their help.
\end{acknowledgments}

\end{document}